\documentclass[12pt,a4,portrait]{article}

\usepackage{latexsym}
\usepackage{hyperref}
\usepackage{epsfig}
\usepackage{amssymb}
\usepackage{psfrag}
\newcommand{\be}{\begin{equation}}
\newcommand{\ee}{\end{equation}}
\newcommand{\beq}{\begin{eqnarray}}
\newcommand{\eeq}{\end{eqnarray}}
\newcommand{\bed}{\begin{displaymath}}
\newcommand{\eed}{\end{displaymath}}
\newcommand{\bc}{\begin{center}}
\newcommand{\ec}{\end{center}}
\newcommand{\bi}{\begin{itemize}}
\newcommand{\ei}{\end{itemize}}
\newcommand{\bn}{\begin{enumerate}}
\newcommand{\en}{\end{enumerate}}

\newcommand{\rw}{ {\rm w} }

\begin{document}

\title{Scalar-tensor cosmologies with a potential\\in the general relativity limit:
time evolution}
\author{Laur J\"arv\thanks{laur.jarv@ut.ee}, Piret Kuusk\thanks{piret@fi.tartu.ee} \\
 Institute of Physics, University of Tartu, Riia 142, Tartu 51014, Estonia \\ \\
 Margus Saal\thanks{margus@fi.tartu.ee} \\
 Tartu Observatory, T\~oravere 61602, Estonia}
\date{}
\maketitle

\begin{abstract}
We consider Friedmann-Lema\^{\i}tre-Robertson-Walker flat cosmological models 
in the framework of general Jordan frame scalar-tensor theories of gravity with arbitrary 
coupling function and potential. 
For the era when the cosmological energy density of the scalar potential dominates 
over the energy density of ordinary matter,
we use a nonlinear approximation of the decoupled scalar field equation 
for the regime close to the so-called limit of general relativity
where the local weak field constraints are satisfied.  
We give the solutions in cosmological time 
with a particular attention to the classes of models 
asymptotically
approaching general relativity.
The latter can be subsumed under two types: (i) 
exponential convergence, and (ii) damped oscillations around
general relativity.
As an illustration we present an example of 
oscillating dark energy.

\end{abstract}


\section{Introduction}
In our recent paper \cite{meie7} we considered the Friedmann-Lema\^{\i}tre-Robertson-Walker (FLRW) flat
cosmological models in the framework of general scalar-tensor theories of gravity (STG) 
with arbitrary coupling functions, set in the Jordan frame, in the era when 
the energy density of the scalar potential dominates over the energy density of ordinary matter.
We used a nonlinear approximation of the decoupled scalar field equation 
and analyzed the corresponding phase portraits in the neighbourhood 
of the point in the phase space which corresponds to general relativity (GR). 
We argued that the topology 
of phase portraits of approximate nonlinear scalar field equation is
similar to the topology of phase portraits of the exact equation. 
In the present paper we 
extend the study to the 
evolution in the cosmological time.        
       
According to contemporary cosmological precision data,
the expansion of the universe has been accelerating for the last few billion years. The
phenomenon can be accommodated in a specific FLRW model with a suitable cosmological constant 
$\Lambda$  which is equivalent to an additional matter component
with an equation of state (EoS) $p = {\rm w} \rho$, ${\rm w} = -1$.     
However, several recent data analyses 
indicate that the dark energy 
(effective) barotropic index ${\rm w}$ is actually
changing in the cosmological time \cite{Huang:2009rf, cai}. In particular, there may 
have been periods with dark energy in the phantom regime, ${\rm w} < -1$, which has inspired  
investigations of cosmological models allowing to cross the phantom divide value ${\rm w} = -1$ \cite{phantom_stg}. 
Alternatively, according to some authors \cite{SSS, Guimaraes:2010mw} the
acceleration may already have passed the peak and is currently slowing down, while others have looked for   
the oscillating patterns in 
the dark energy which does not cross the phantom divide \cite{Lazkoz:2010gz}.
Related theoretical models have been also proposed \cite{sadeghi}. 
     
In this paper we present approximate solutions of the STG FLRW equations in the era when
the energy density of the scalar potential dominates 
over the energy density of ordinary matter and demonstrate that depending on the choice of the
model parameters, they include crossing the phantom divide as well as oscillating dark energy.
The plan of the paper is the following.
We begin in Sec.~2 with the general equations of STG FLRW cosmological models and their
nonlinear approximation introduced in our previous paper \cite{meie7}. Sec.~3 
presents the solutions of approximate decoupled scalar field equation and gives an additional hint that
the same types of solutions occur also in the full theory. An example of oscillating effective barotropic index
of the dark energy is given in Sec.~4. Finally, Sec. 5 provides concluding remarks.


\section{Equations}

We consider a general scalar-tensor gravity 
in the Jordan frame
governed by the action functional
\beq \label{jf4da}
S  = \frac{1}{2 \kappa^2} \int d^4 x \sqrt{-g}
        	        \left[ \Psi R(g) - \frac{\omega (\Psi ) }{\Psi}
        		\nabla^{\rho}\Psi \nabla_{\rho}\Psi
                  - 2 \kappa^2 V(\Psi)  \right] 
                  + S_m\,.
\eeq 
Here $\omega(\Psi)$ is a coupling function (we assume $2 \omega (t) + 3 \geq 0 $ to avoid ghosts in the
Einstein frame, see  e.g.  Ref. \cite{polarski}) 
and $V(\Psi)\geq 0$ is a potential, $\nabla_{\mu}$ 
denotes the covariant derivative with respect to the metric 
$g_{\mu\nu}$, $S_m$ is the matter action, and $\kappa^2$ is the non-variable part of
the effective gravitational constant $\frac{\kappa^2}{\Psi}$.
In order to keep the latter positive
we assume that $0 < \Psi < \infty$.

The field equations for the flat FLRW line element 
\be
ds^2=-dt^2 + a(t)^2 \left(dr^2 + r^2 (d\theta ^2 + \sin ^2 \theta d\varphi ^2)\right)
\ee
and barotropic fluid ($p=\rw \rho$) read 
\beq 
\label{00}
H^2 &=& 
- H \frac{\dot \Psi}{\Psi} 
+ \frac{1}{6} \frac{\dot \Psi^2}{\Psi^2} \ \omega(\Psi)
+ \frac{\kappa^2}{\Psi} \frac{\rho}{3} 
+ \frac{\kappa^2}{\Psi} \frac{V(\Psi)}{3} \,, 
\\ \nonumber \\
\label{mn}
2 \dot{ H} + 3 H^2 &=& 
- 2 H \frac{\dot{\Psi}}{\Psi} 
- \frac{1}{2} \frac{\dot{\Psi}^2}{\Psi^2} \ \omega(\Psi) 
- \frac{\ddot{\Psi}}{\Psi} 
- \frac{\kappa^2}{\Psi} \rw \rho
+ \frac{\kappa^2}{\Psi} \ V(\Psi) \,, 
\\ 
\label{deq}
\ddot \Psi &= & 
- 3H \dot \Psi 
- \frac{1}{2\omega(\Psi) + 3} \ \frac{d \omega(\Psi)}{d \Psi} \  \dot {\Psi}^2 
+ \frac{\kappa^2}{2 \omega(\Psi) + 3} (1-3\rw)\rho \nonumber\\
&& \qquad + \frac{2 \kappa^2}{2 \omega(\Psi) + 3} \ \left[ 2V(\Psi) - \Psi \ 
\frac{d V(\Psi)}{d \Psi}\right] \, ,
\eeq 
where $H \equiv \dot{a} / a$.

We can express the Hubble parameter $H$ as a function of $\Psi$ by solving the Friedmann equation (\ref{00}) algebraically,
\be  \label{H_eq}
H = -\frac{\dot\Psi}{2\Psi} \pm \sqrt{(2\omega(\Psi)+3)\frac{\dot\Psi^2}{12 \Psi^2} + \frac{\kappa^2 (\rho + V(\Psi))}{3 \Psi}} \,.
\ee 
Upon introducing the notation
\be
A(\Psi) \equiv \frac{d}{d\Psi} \left(\frac{1}{2\omega(\Psi)+3} \right)\,, \qquad
W(\Psi) \equiv 2 \kappa^2  \left(2V(\Psi)-\frac{dV(\Psi)}{d\Psi}\Psi \right)\,
\ee
and substituting Eq. (\ref{H_eq}) in Eq. (\ref{deq}),
we get a decoupled equation for the scalar field
\beq 
\label{ddotpsi}  
\ddot{\Psi} &=& \left( \frac{3}{2\Psi} + \frac{1}{2} A(\Psi) (2\omega(\Psi)+3)\right) \dot{\Psi}^2
     + \frac{\kappa^2}{2 \omega(\Psi) + 3} (1-3\rw)\rho \nonumber \\
    & &\pm  \frac{1}{2\Psi} \sqrt{3( 2\omega(\Psi) + 3) \dot{\Psi}^2   
+ 12 \kappa^2 \Psi V(\Psi)} \ \dot{\Psi}   
 +\frac{W(\Psi)}{2\omega(\Psi)+3}  \,.
\eeq 

In the limit  $\frac{1}{(2\omega(\Psi)+3)} \rightarrow 0$, $\dot\Psi \neq 0$ the system 
faces a spacetime curvature singularity, since $H$ diverges, and likewise behaves $\ddot\Psi$.
At first the limit (a) $\frac{1}{(2\omega(\Psi)+3)} \rightarrow 0$, (b) $\dot\Psi \rightarrow 0$ 
seems only slightly less mathematically precarious
for the equations are left just indeterminate (contain  terms $\frac{0}{0}$) . 
Yet the latter situation is of particular physical importance, as 
the experiments in the Solar System (where matter density dominates over the scalar potential), 
i.e. the limits of observed values of the parametrized post-Newtonian (PPN) parameters  
and the time variation of the gravitational constant \cite{weakfield},
\beq
8 \pi G &=& \frac{\kappa^2}{\Psi} \frac{2\omega+4}{2\omega+3} \\
\label{beta_exp}
\beta - 1 &\equiv& \frac{\kappa^2}{G} \frac{\frac{d \omega}{d \Psi}}{(2 \omega +3)^2 (2 \omega+4)} 
\lesssim \, 10^{-4} \\
\label{gamma_exp}
\gamma -1 &\equiv& - \frac{1}{\omega + 2}  \lesssim \ 10^{-5} \\
\label{Gdot_exp}
\frac{\dot G}{G} &\equiv& - \dot{\Psi} \frac{2\omega+3}{2\omega+4} 
\left( G + \frac{2 \frac{d \omega}{d \Psi}}{(2\omega + 3)^2} \right) \lesssim \ 10^{-13} \ {\rm yr^{-1}} \,,
\eeq
suggest the present cosmological background value of the scalar field to be very close to the limit 
(a)-(b). 
As in this limit the STG PPN parameters coincide with those of general relativity we may tentatively 
call (a)-(b) `the limit of general relativity.'

Let us define $\Psi_{\star}$ by $\frac{1}{2\omega(\Psi_{\star})+3} =0$.
In our previous papers \cite{meie7, meie4, meie5} we studied the limit 
(a)-(b) 
with the simplifying assumptions (c) $A_{\star} \equiv A(\Psi_{\star}) \neq 0$ and 
(d) $\frac{1}{2\omega+3}$ is differentiable at $\Psi_{\star}$, which enabled to
Taylor expand the functions in Eq. (\ref{ddotpsi}) 
and find analytic solutions in the phase space for the resulting approximate equation.  
The outcome was that the solutions are well behaved in this limit, motivating the 
inclusion of $(\Psi=\Psi_{\star}, \dot\Psi=0$) as a boundary point to the open domain of 
definition of Eq. (\ref{ddotpsi}). 
Moreover, it was possible to identify a wide class of STGs 
where the FRLW cosmological dynamics 
spontaneously draws the scalar field to this limit, i.e.  into agreement with 
current local weak field observations in the Solar System.

The aim of the present paper is to complement our previous study \cite{meie7} of the cosmological
epoch when the energy density of the scalar potential dominates over the energy density of
the ordinary matter 
by considering the limit (a)-(d) and finding analytic solutions 
not in phase space variables but explicitly in time,
in order to facilitate the assessment of their cosmological viability.
Let us begin by introducing  $\Psi(t) = \Psi_{\star} + x(t)$,
and assuming that $x(t)$, $\dot{x}(t)$ are both first order small.
Then
\be
\frac{1}{2\omega(\Psi (t))+3} = \frac{1}{2\omega(\Psi_{\star})+3}+ A_{\star} x(t)+ ...\approx A_{\star} x(t) \,.    
\ee
By denoting $W_{\star} \equiv W(\Psi_{\star})$ and
\be
\label{def}
 C_1 \equiv \pm \sqrt{\frac{3 \kappa^2 V(\Psi_{\star})}{\Psi_{\star}}} \,, \qquad
C_2 \equiv A_{\star}\, W_{\star} \,, 
\ee
the decoupled scalar field equation (\ref{ddotpsi}) with $\rho = 0$ is approximated by
\be
\label{mlin_eq}
{\ddot x} + C_1 \ {\dot x} - C_2 \ x =  \frac{{\dot x}^2}{2x} \,.
\ee
We will investigate its solutions in the next section.  

Analogously we may express
\beq 
\label{approx_H}
H &=& \frac{C_1}{3} - \frac{1}{2\Psi_{\star}} \dot{x}
+ \frac{1}{2C_1 \Psi_{\star}} 
\left(\frac{C_1^2}{3} - \frac{C_2}{2 A_{\star} \Psi_{\star}} \right) x 
+ \frac{1}{8 C_1 \Psi_{\star}A_{\star}} \ \frac{\dot{x}^2}{x} + \ldots \,, \\
\dot{H} &=& \frac{2C_1}{3 \Psi_{\star}} \dot{x} - \frac{C_2}{2 \Psi_{\star}} x 
-\frac{1+A_{\star} \Psi_{\star}}{4 A_{\star} \Psi_{\star}^2} \frac{\dot{x}^2}{x}  + \ldots \,, \\
\label{wii}
\rw_{\rm eff} &\equiv & -1 - \frac{2 {\dot H}}{3H^2} = 
-1 + \frac{1}{C_1^2 \Psi_{\star}} \left[\frac{3}{2} \left(1 + \frac{1}{\Psi_{\star}A_{\star}}\right) 
\frac{\dot{x}^2}{x} - 4 C_1 \dot{x} + 3 C_2 x  \right]  + \ldots \,.
\eeq
A necessary condition for crossing the phantom divide $\rw_{\rm eff} = -1 $ is vanishing of 
the second term  in Eq. (\ref{wii}). This occurs if
 $\dot{x} $ equals to a solution of the corresponding quadratic equation
\be \label{K_pm}
 \dot{x} = \frac{A_{\star} \Psi_{\star}}{3 (1 + A_{\star} \Psi_{\star})} \left[ C_1 \pm
\sqrt{D } \right] x \equiv K_\pm x\,, \qquad
D \equiv C_1^2 - \frac{9 C_2}{8} \left(1+ \frac{1}{A_{\star} \Psi_{\star}} \right)\,.
\ee
Here we can see a condition on the constants $ A_{\star}, \Psi_{\star}, C_1,C_2 $ which characterize
the theory: $ K_\pm$ must be real numbers, i.e. $D \geq 0 $. 
Note that on the plane $(x, \dot{x}) $ the phantom divide (\ref{K_pm})  consists of two straight lines 
crossing at the origin $x = 0$, $\dot{x} = 0$.

The PPN parameters (\ref{beta_exp}), (\ref{gamma_exp}) and $\frac{{\dot G}}{G}$ (\ref{Gdot_exp}) 
are approximated by 
\beq
\beta - 1 &=& - \frac{1}{2} A_{\star} \Psi_{\star} x + \ldots \,, \label{beta} \\
\gamma -1 &=& - 2 A_{\star} x + \ldots \,, \label{gamma} \\
\frac{\dot G}{G} &=& - \frac{1 - A_{\star} \Psi_{\star}}{\Psi_{\star}} \dot{x} + \ldots \,.
\eeq
Strictly speaking, for constraining cosmological models using PPN type observations requires solutions with
$\rho \not= 0$.  In the following we will consider cosmological solutions with $\rho = 0, V(\Psi) \not= 0$ 
for the late time universe, but since the matter density in the Solar System exceeds considerably the 
cosmological  matter density 
local gravitational experiments are still described by ordinary matter dominated theory 
and the PPN constraints must be satisfied.  

\section{Solutions}
\label{solutions}

The general solution of Eq. (\ref{mlin_eq}) reads
\be
\label{mlin_sol}
\pm x(t) = e^{-C_1t} \left[ M_1 e^{\frac{1}{2}t \sqrt{C_1^2 + 2 C_2}} -  
M_2 e^{-\frac{1}{2}t \sqrt{C_1^2 + 2 C_2}} \right]^2 \,,
\ee
where $M_1$ and $M_2$ are arbitrary constants of integration. 
These solutions fall into
three classes, depending on the sign of constant $C \equiv C_1^2 + 2 C_2$.
A detailed classification and characterization of the solutions,
including the phase portraits was given in Ref. \cite{meie7},
here we focus only on these solutions which approach the GR limit 
asymptotically in time.

\subsection{``Exponential solutions''} 

In the case $C > 0$ solutions read
\be 
\label{exp}
\pm x = e^{-C_1t}\left[M_1 e^{\frac{1}{2}t \sqrt{C}}  -  
M_2 e^{-\frac{1}{2}t \sqrt{C}} \right]^2 \,.
\ee
If $C_1>0$ and $C_2<0$ the 
solutions exponentially converge to the GR limit, behaving as
\be
\pm x |_{t \rightarrow \infty} =  e^{-(C_1-\sqrt{C})(t - t_1)} \,
\ee
Here we have denoted the constant of integration as $M_1 \equiv  e^{-\frac{1}{2}t_1 (\sqrt{C} - C_1)} $
for some arbitrary moment $t_1$.  
All solutions satisfy an asymptotic condition
\be 
\frac{{\dot x}}{x}|_{t \rightarrow \infty} = \frac{\sqrt{C} - C_1}{\sqrt{C}} \,.
\ee 
The Hubble parameter reads
\be  
H|_{t \rightarrow \infty} = \frac{C_1}{3} \pm \frac{e^{-(C_1-\sqrt{C})(t - t_1)}}{2 \Psi_{\star}} 
\left[\frac{C_1 - \sqrt{C}}{2A_{\star} \Psi_{\star}} + \frac{4}{3} C_1 -  \sqrt{C}\right]  \,  
\ee
and effective barotropic index (\ref{wii}) can be calculated
\be \label{wii_exp}
{\rm w_{eff}}|_{t \rightarrow \infty} = -1 \pm \frac{e^{-(C_1-\sqrt{C})(t - t_1)}}{ C_1^2  \Psi_{\star}}  \left[
\frac{3 (C_1 - \sqrt{C})^2 }{2 A_{\star} \Psi_{\star}} + 4 C_1^2 + 3 C - 7 C_1 \sqrt{C}) \right] 
 \,.
\ee
Now we can determine whether a model in the theory characterized by distinct parameters 
($C_1, C \equiv C_1^2 + 2C_2, A_{\star}$)
approaches the de Sitter spacetime from quintessence side (${\rm w_{eff}} > -1 $) or from the phantom  
side (${\rm w_{eff}} < -1$).
Approximate expressions of the PPN parameters (\ref{beta}), (\ref{gamma})
indicate that they approach the GR values $\beta = 1 $, $\gamma = 1 $ exponentially.

Solutions (\ref{exp}) may have interesting features at certain finite moments of time. Firstly, if the theory allows phantom divide, i.e. if $K_pm$ in Eq. (\ref{K_pm}) are real numbers,
then solutions (\ref{exp}) may cross the phantom 
divide no more than at two moments $t_\pm$:
 \be
t_\pm = \ln\frac{M_2 (C_1 + K_\pm + \sqrt{C})}{M_1 (C_1 + K_\pm - \sqrt{C})} \,.
\ee
Secondly, at finite moments $t_b$ some solutions can achieve  $x(t_b)=0$, ${\dot x}(t_b) = 0$  depending on values of 
integration constants $M_1$, $M_2$. The latter ones can be given in terms of initial values  e.g. at $t = 0$,
$x(0) = x_0$, ${\dot x}(0) = {\dot x}_0$. 
Indeed, there is a moment $t_b$ such that $x(t_b)=0$, ${\dot x}(t_b) = 0$, namely
\be
t_b  = \frac{1}{C} \ln \frac{M_2}{M_1} = \frac{1}{C}\ln\left[
\frac{{\dot x}_0 +  x_0(C_1 - C)}{{\dot x}_0 + x_0(C_1 + C)} \right]\,.
\ee
Phase trajectories have a vertical slope there and can be described as "turning back" if we
consider solutions (\ref{exp}) with only one sign (+ or -). The region of initial values for trajectories with
this property can be found from the condition
\be
\frac{{\dot x}_0 +  x_0(C_1 - C)}{{\dot x}_0 + x_0(C_1 + C)} >0 \,.
\ee
On the half-plane $x>0$ it is a region between two rays ${\dot x}_0 =  x_0(C - C_1)$ and 
${\dot x}_0 = - x_0(C + C_1)$.
The corresponding region on the other half-plane $x<0$ is its mirror image with respect to the origin
 $x=0$, ${\dot x}=0$.

\subsection{``Linear exponential solutions''}

In the case $C=0$ the solutions read
\be
\label{quadr}
\pm x = e^{-C_1t}\left[e^{ \frac{1}{2} C_1 t_1} t -  M_2 \right]^2 \, 
\ee
with $M_1 \equiv e^{ \frac{1}{2} C_1 t_1} $.
If $C_1>0$ the 
solutions exponentially converge to the GR limit, behaving as
\be
\pm x|_{t \rightarrow \infty} =  t^2 e^{-C_1 (t - t_1)} \,,
\ee 
the Hubble parameter reads
\be
H|_{t \rightarrow \infty} = \frac{C_1}{3} \pm \frac{C_1 }{\Psi_{\star}} \left[ \frac{2}{3} +
\frac{1}{4 A_{\star} \Psi_{\star}} \right] t^2 e^{-C_1 (t - t_1)}  
\ee
and ${\rm w_{eff}}$ behaves as
\be
{\rm w_{eff}}|_{t \rightarrow \infty} = -1 \pm \frac{M_1^2}{\Psi_{\star}} \left[\frac{1}{A_{\star}\Psi_{\star}}
+ 4   \right] t^2 e^{-C_1 (t- t_1)} \,.
\ee
However, this case is rather finetuned by the condition $C_1^2 = - 2 C_2 $.

\subsection{``Oscillating solutions''}

In the case $C<0$ the solutions read
\be 
\label{sin}
\pm x = e^{-C_1t} \left[N_1 \sin(\frac{1}{2}t \sqrt{|C|}) - N_2 \cos (\frac{1}{2}t \sqrt{|C|}) \right]^2 \,,
\ee
where $N_1$, $N_2$ are integration constants.
In terms of the phase space $(x, {\dot x})$ solutions do not have a definite slope 
at approaching asymptotically
($t \rightarrow \infty$) to $x=0$, ${\dot x}=0$. 
As they spiral to it at $t \rightarrow \infty$ their phase trajectories cross 
this point infinitely many times at moments $t_n$, 
\be
t_n = \frac{2}{\sqrt{|C|}} \left[\arctan \frac{N_2}{N_1} + n \pi \right] \,.
\ee
 The spiral, however, 
must lie in one half-plane of domain of definition, either $x > 0$ or $x < 0$. 
Approximate expressions of the PPN parameters (\ref{beta}), (\ref{gamma}) now
reveal damped oscillatory behaviour around the GR values.  

If the theory allows crossing the phantom divide, i.e. if $K_\pm$ in Eq. (\ref{K_pm}) are real, then the 
possible moments $t_{\pm n}$ of crossing occur on each winding of the spiral:
\be 
t_{\pm n} = \frac{2}{\sqrt{|C|}} \left[ \arctan \frac{(C_1 + K_\pm) N_2 + 
\sqrt{|C|} N_1}{(C_1 + K_\pm) N_1 - \sqrt{|C|} N_2} + n \pi \right] \,.
\ee  
If $K_\pm$ is imaginary, then the effective barotropic index ${\rm w_{eff}}$ stays below or above $-1$.

\subsection{A hint for the full theory}

We have considered solutions (\ref{mlin_sol}) of an approximate equation (\ref{mlin_eq}) for
the scalar field. Among them the most peculiar seem to be ``oscillating solutions'' and a question 
arises whether this type of solutions really occur in the full theory (\ref{00})--(\ref{deq}), 
or perhaps they are only an artefact of approximation. 
In our previous paper we argued on the basis of the 
phase space that the topology of trajectories in the nonlinear approximation is representative of those 
of the full system, and thus the correspondence holds \cite{meie7}.

Here we supplement this discussion by a specific yet rather unphysical model, where an additional indication can be found that 
such solutions indeed exist in the full theory.
Let us choose the two functional degrees of freedom to be
\be 
\omega (\Psi) = \frac{3 \Psi}{2(1-\Psi)} \,, \qquad \kappa^2 V(\Psi) = a (1 - \Psi), \quad
 a = {\rm const} > 0 \,.
\ee 
In the corresponding approximate theory we have $\Psi \in (0, 1]  $,
$\Psi_{\star} =1 $,  $A_{\star} = - \frac{1}{3} $, $C_1 = 0 $,
$C_2 = - \frac{2a}{3} $, $C = -\frac{4a}{3} $, and thus  the approximate solutions
should belong to the oscillatory type.

In the exact theory the Friedmann constraint (\ref{00}) reads
\be   \label{naide} 
H^2  + H \frac{\dot \Psi}{\Psi} - \frac{{\dot \Psi}^2}{4 \Psi (1 - \Psi)} - \frac{a(1-\Psi)}{3\Psi} = 0 \,.
\ee
Considering a subspace $H = {\rm const}$ in the phase space  ($H, \Psi, {\dot \Psi}$), 
Eq. (\ref{naide}) is a second order algebraic equation for the phase trajectories 
\be  
\left(H^2 + \frac{a}{3}\right) \Psi^2 + H {\dot \Psi} \Psi + \frac{1}{4} {\dot \Psi}^2 
- \left(H^2 + \frac{2a}{3}\right) \Psi - H {\dot \Psi} + \frac{a}{3} = 0
\ee
and their
topology (ellipses or hyperbolas) is given by the coefficients of the second order terms:
\be
H^2 - 4 (H^2 + \frac{a}{3}) \frac{1}{4} = - \frac{a}{3} < 0 \,.
\ee
It follows that the phase trajectories must indeed be ellipses, i.e. we have oscillations.

\section{An example}

To illustrate the behaviour of oscillating solutions, let us specify the coupling function and potential as
\beq
\omega(\Psi) = \frac{\Psi}{2 (1- \Psi)}
\,, \qquad
\kappa^2 V(\Psi) = V_{0} e^{3(1-\Psi)} \,.
\eeq
Here $\Psi \in (0, 1]$ and the `GR point' is at $\Psi_{\star} =1$.
It makes sense to measure the oscillations in the units of the analogue of Hubble time, $T = H_{\star} \, t = \frac{C_{1}}{3} \,t $. 
Using prime to denote derivation with respect to $T$,
the integration constants $N_{1}$ and $N_{2}$ are related to the initial conditions of
$x_{0} = x(0)$ and $x'_{0} = x'(0)$ as
\beq
\label{example_N}
N_{1} =  \frac{x'_{0} + x_{0} \, C_{1}}{\sqrt{-x_{0}} \, \sqrt{|C|}}\, ,  \qquad N_{2} = \sqrt{-x_{0}} \,.
\eeq
To satisfy the current weak field constraints (\ref{beta_exp})--(\ref{Gdot_exp})  
we choose the initial conditions  
\beq
\label{example_x0}
x_{0} = -0.000025 \, , \qquad x'_{0}=  0.0025 \,.
\eeq
Substituting the solution (\ref{sin}) and constants (\ref{example_N}), (\ref{example_x0})
into the expansion (\ref{wii}) we obtain the evolution of ${\rm w_{eff}}$, which is plotted on Fig. 1.
Here we see the effective barotropic index going through the
phantom divide line in a damped oscillatory manner. 
We also note that the period of oscillations is comparable to the Hubble time, 
roughly
the same order of magnitude as the age of the Universe.
This suggests a possible scenario for the universe, where during the matter domination era the scalar field has  already relaxed close to the GR limit \cite{meie5}, while later when the cosmological energy density of the potential
becomes more significant, we should observe the effects of
slowly oscillating dark energy.

\begin{figure}[t]
\begin{center}
\psfrag{weff}{$\rm{w}_{eff}$}
\psfrag{T}{{$T$}}
\includegraphics[width=8cm,angle=-90]{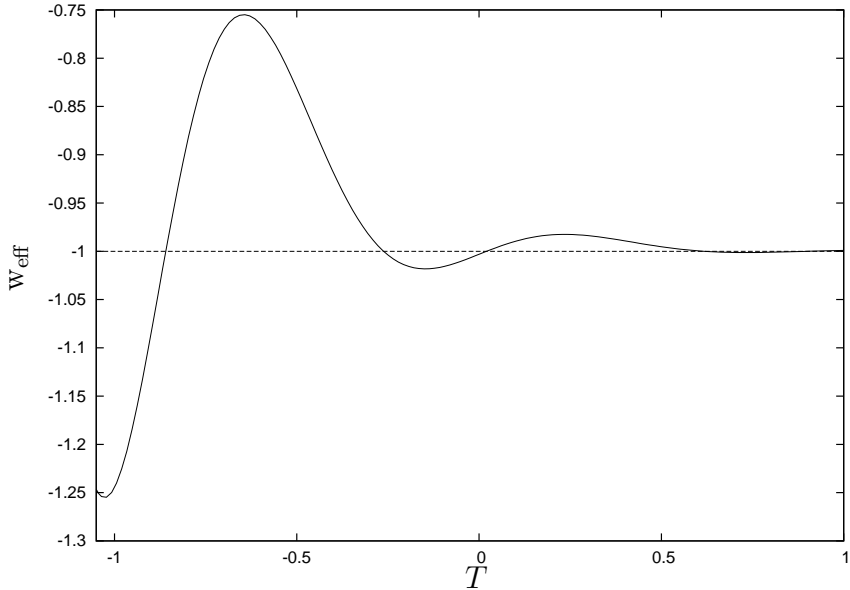}
\end{center}
\caption
{An example of oscillating dark energy described in Sec. 4.}
\label{per_weff}
\end{figure}

\section{Concluding remarks}

We have proposed and solved a nonlinear approximate equation (\ref{mlin_eq}) for decoupled scalar
field in the framework of STG FLRW cosmological models 
describing the era when the energy density 
of the scalar potential dominates over the energy density of
the ordinary matter and the universe has evolved close to the limit of general relativity.
The behaviour of solutions which approach general relativity
can be classified under 
two characteristic types: (i) ``exponential'' or ``linear exponential'' convergence, 
and (ii) damped ``oscillations'' around general relativity.
These solutions also determine the evolution of cosmological and PPN parameters. 
Depending on the model, exponential solutions may cross the phantom divide line at most 
twice before approaching ${\rm w_{eff}=-1}$ from either above or below. 
In the oscillating type the dark energy effective barotropic index oscillates either in the 
quintessence regime, phantom regime, or crossing the phantom divide line once or twice during 
each period. 
The latter case is illustrated by an explicit example.


\bigskip
{\bf Acknowledgments}
\smallskip

This work was supported by the Estonian Science Foundation Grant No. 7185 and by 
Estonian Ministry for Education and Science Support Grant No. SF0180013s07. 
M.S. also acknowledges the Estonian Science Foundation Postdoctoral research Grant No. JD131.

\end{document}